\documentclass[apj]{emulateapj}
\usepackage{apjfonts} 

\newcommand{\n}{\nodata}

\makeindex
\citeindextrue

\shorttitle{The Fifth VLBA Calibrator Survey}
\shortauthors{Kovalev et al.}
\received{2006 July 23}
\revised{2006 November 15}
\accepted{2006 November 19}
\journalinfo{Astronomical~journal}
\submitted{}

\begin{document}

\title{The Fifth VLBA Calibrator Survey: VCS5}

\author{Y. Y. Kovalev\altaffilmark{\dag}}
\affil{National Radio Astronomy Observatory,
       P.O.~Box 2, Green Bank, WV~24944, USA;}
\affil{Max-Planck-Institut f\"ur Radioastronomie,
       Auf dem H\"ugel 69, 53121 Bonn, Germany;}
\affil{Astro Space Center of Lebedev Physical Institute,
       Profsoyuznaya 84/32, 117997 Moscow, Russia}
\email{ykovalev@mpifr-bonn.mpg.de}
\author{L. Petrov\altaffilmark{\ddag}}
\affil{Mizusawa Astrogeodynamics Observatory, NAOJ, Mizusawa 023-0861, Japan}
\email{Leonid.Petrov@lpetrov.net}
\author{E. B. Fomalont}
\affil{National Radio Astronomy Observatory,
       520 Edgemont Road, Charlottesville, VA~22903--2475, USA}
\email{efomalon@nrao.edu}
\author{D. Gordon}
\affil{Raytheon/NASA GSFC, Code 698, Greenbelt, MD 20771, USA}
\email{dgg@leo.gsfc.nasa.gov}
\altaffiltext{\dag}{Jansky Fellow, National Radio Astronomy Observatory}
\altaffiltext{\ddag}{on leave from NVI, Inc. / NASA GSFC}

\begin{abstract}

This paper presents the fifth part of the Very Long Baseline Array
(VLBA) Calibrator Survey (VCS), containing 569 sources not observed
previously with very long baseline interferometry in geodetic or 
absolute astrometry programs. This campaign has two goals: (i) to
observe additional sources which, together with previous survey results,
form a complete sample, (ii) to find new strong sources suitable as
phase calibrators. This VCS extension was based on three 24-hour VLBA
observing sessions in 2005. It detected almost all extragalactic
flat-spectrum sources with correlated flux density greater than 200~mJy
at 8.6 GHz above declination $-30\degr$ which were not observed
previously. Source positions with milliarcsecond accuracy were
derived from astrometric analysis of ionosphere-free combinations of
group delays determined from the 2.3~GHz and 8.6~GHz frequency bands. The VCS5
catalog of source positions, plots of correlated flux density versus
projected baseline length, contour plots and FITS files of naturally
weighted CLEAN images, as well as calibrated visibility function files
are available on the Web at \url{http://vlbi.gsfc.nasa.gov/vcs5}.

\end{abstract}


\section{Introduction}
\label{s:introduction}

    This work is a continuation of the survey search for bright
compact radio sources. Several major applications require an
extended list of sources with positions known at the nanoradian
level: geodetic observations including space navigation; very
long baseline interferometry (VLBI) phase-referencing of weak
targets, and differential astrometry. For satisfying the needs of
these applications, 878 sources were observed under various
geodetic and astrometric programs from 1979 thorough 2002, and
over 80\,\% of them were detected. Results of these observations
were presented in the catalog ICRF-Ext.2
\citep{icrf-ext2-2004} that contains positions of 776 sources.
   Additionally, 2952 flat-spectrum sources were observed in nineteen
24-hour sessions from 1994 through 2005 in the Very Long Baseline Array
(VLBA) Calibrator Survey (VCS) program. The positions of 2505 sources
were determined from the observations of the VCS project: VCS1
\citep{VCS1}, VCS2 \citep{VCS2}, VCS3 \citep{VCS3}, and VCS4
\citep{VCS4}. Since 364 sources are listed in both the ICRF-Ext.2 and
the VCS catalogs, the total number of sources for which positions were
determined with VLBI in IVS (International VLBI Service for astrometry
and geodesy) and VCS1 to VCS4 experiments is 2917. Among them, 2468
sources, or 85\,\%, are considered acceptable calibrators: having at
least eight successful observations at both X band (central frequency
8.6~GHz) and S band (central frequency 2.3~GHz), and the
semi-major axis of the error ellipse of their coordinates being less
than 25~nrad ($\approx$~5~mas). When observations from both the geodetic
programs and VCS1--4 are combined, the overall catalog provides fairly
good sky coverage. The probability of finding a calibrator within
4\degr\ of any target north of declination $-40\degr$ is 98.1\,\%.

In this paper we present an extension to the VCS catalogs, called the
VCS5 catalog. It concentrates on the brightest flat-spectrum sources
north of declination $-30\degr$ not previously observed with VLBI under
geodetic and absolute astrometry programs. VCS5 is different from the
previous campaigns since its prime goal is to collect data needed for
astrophysical analysis of active galactic nuclei (see
\S\ref{s:objectives}).

Since the observations, calibration, astrometric solutions and imaging
are similar to that of VCS1--4, most of the details are described by
\cite{VCS1} and \cite{VCS3} and will not be repeated here.
In \S\ref{s:objectives} we discuss
scientific objectives for the VCS5 survey. In \S\ref{s:selection} we
describe the strategy for selecting the 675 candidate sources observed in
three 24-hour VCS5 sessions with the VLBA based on analysis of the
available multi-frequency non-VLBI continuum radio measurements. The
same strategy was successfully applied by us earlier to select one
hundred objects with the strongest estimated flux density at 8.6~GHz in
the framework of the VCS4 survey. Sixty seven out of these one hundred
VCS4 candidates showed X~band correlated flux density greater than
0.2~Jy \citep{VCS4}. In \S\ref{s:obs_anal} we briefly outline the
observations and data processing. We present the VCS5 catalog in
\S\ref{s:catalog}, and summarize our results in \S\ref{s:summary}.

\section{Scientific objectives for the VCS5 survey}
\label{s:objectives}

     There are two main scientific objectives for the VCS5 survey. We
would like to perform statistical analysis of physical properties of a
deep sample of compact AGNs on the basis of milliarcsecond scale images
measured simultaneously at S~band and X~band with the VLBA. A cursory
analysis of the sample of the 2917 VCS/ICRF sources observed with the
VLBA in the S/X mode revealed that it is nearly complete down to
0.5~Jy but becomes incomplete at lower flux densities. As a result,
possible usage of this largest collection of VLBI data for statistical
analysis of properties of active galactic nuclei at milliarcsecond
scales is limited. The first goal of the current VCS5 project is to
observe the remaining bright sources with expected correlated flux
densities in the range 200~mJy to 600~mJy to create a statistically
complete sample  of extragalactic flat-spectrum radio sources with
integrated flux density  at milliarcsecond scales greater than 200~mJy
at X~band. This will make results of the VCS survey significantly more
useful for astrophysical applications. The uniformity of VCS data
reduction as well as the completeness and homogeneity of the source
sample will guarantee robust results from further statistical studies.

The second goal is to find more compact sources and to measure their
positions precisely for use in geodetic applications including space
navigation, VLBI phase referencing of weak targets, and differential
astrometry. Many applications prefer a more distant bright calibrator to
a near-by but weaker calibrator, since more time can be spent on the
target. The VCS5 observations cover almost all remaining {\em bright}
calibrators with correlated flux density at or greater that 200~mJy.

\section{Source selection}
\label{s:selection}

Our source selection goal was to find all flat-spectrum radio sources
brighter than 0.2~Jy at 8.6~GHz that were missing from the VCS1 to VCS4 and
ICRF-Ext2 catalogs. We define flat radio spectrum as having a spectral
index $\alpha>-0.5$ ($S\sim\nu^\alpha$). To compile a list of missing
objects, we first selected all sources from the NVSS catalog
\citep{NVSS} with flux density at 1.4 GHz $S>50$~mJy, declination
$\delta>-30\degr$, a Galactic latitude $|b|>1\fdg5$, and not identified
with Galactic objects. Since the NVSS catalog is more than 99\,\%
complete for flux density $S>50$~mJy, it is unlikely that sources with
highly inverted spectra and flux density $S>200$~mJy at 8.6~GHz will
have been missed.

\begin{figure}[t]
\begin{center}
\resizebox{1.0\hsize}{!}
{
  \includegraphics[trim=0cm -0.5cm 0cm 0cm]{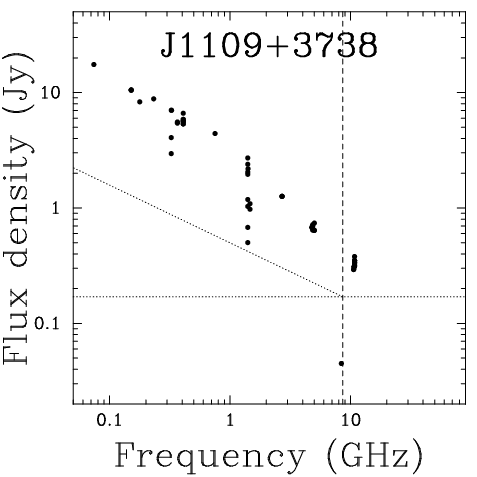}
  \includegraphics[trim=0cm -0.5cm 0cm 0cm]{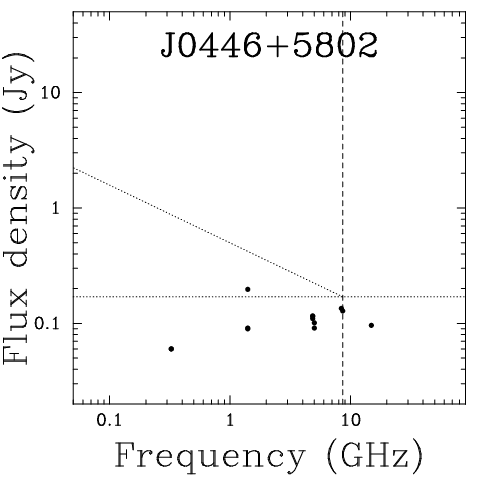}
}
\resizebox{1.0\hsize}{!}
{
  \includegraphics[trim=0cm 0cm 0cm 0cm]{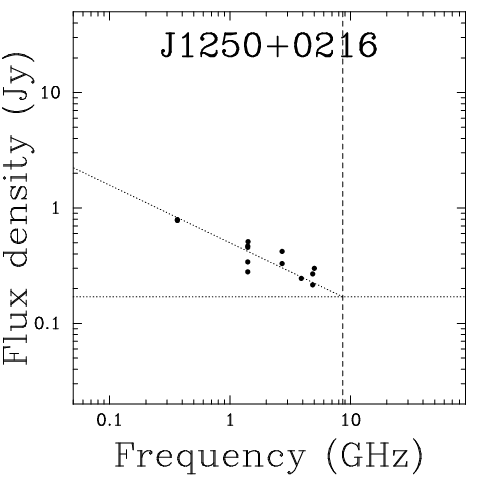}
  \includegraphics[trim=0cm 0cm 0cm 0cm]{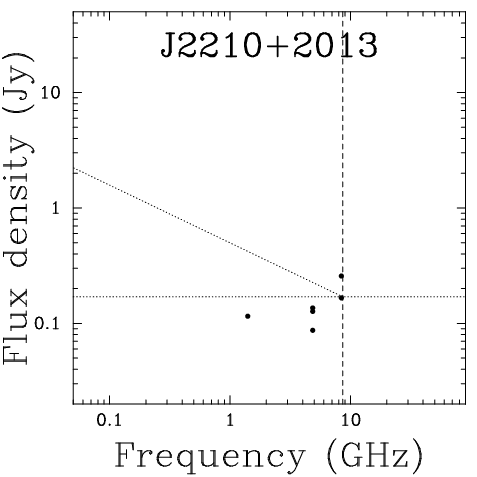}
}
\end{center}
\caption{
Illustration of the candidates selection procedure. The data from all
available publications were located through the CATS database.
The dotted falling lines represent $\alpha=-0.5$.
J1109+3738 was not selected since its integrated spectral
index was less than $-0.5$. 
J0446+5802 was not selected since its flux density extrapolated to
8.6~GHz was less than 170~mJy. J1250+0216 and J2210+2013 satisfied all
the selection criteria mentioned in \S~\ref{s:selection} and were
selected. Their $uv$ and image data are presented in
Figure~\ref{f:images}.
\label{f:selection}
}
\end{figure}

We then searched the CATS database \citep{CATS} containing almost all
radio catalogs to find flux density measurements at other radio
frequencies for the selected NVSS sources.
These data were supplemented by results of the 1--22 GHz
instantaneous broad-band spectra measurements of $\sim$3000
extragalactic flat-spectrum radio sources which we performed at the
transit mode 600~m ring radio telescope \mbox{RATAN-600} of the Russian
Academy of Sciences \citep[see, e.g.,][]{Kovalev_etal99}. The collected
data were then analyzed semi-automatically, and bad data points, wrong
identifications, multiple data points corresponding to different
components of the same extended object were flagged. We found
that we could compile a complete sample of sources with total flux
density spectrum
flatter than $\alpha=-0.5$, and with estimated total flux density of
$S>170$~mJy at 8.6~GHz. In this complete sample were 675 candidates
not previously observed in geodetic VLBI mode, and these are the sources
selected for VCS5 observations.  Figure~\ref{f:selection} presents
examples of plots of the total flux density spectra collected by the
CATS database which we used for source selection.

Our analysis of the multi-frequency catalogs and RATAN observations used
for selection indicates that we have found almost all of the sources
with spectral index greater than $-0.5$ and estimated total flux density
at 8.6~GHz $S>170$~mJy. It is based on the fact that many used catalogs
including NVSS \citep{NVSS}, FIRST \citep{FIRST}, 87GB \citep{87GB}, GB6
\citep{GB6}, CLASS \citep{Myers_etal03}, JVAS \citep{JVAS1,JVAS2,JVAS3},
PMN \citep{Wright_etal94,Griffith_etal95,Wright_etal96}, and PKSCAT90
\citep{pkscat90}, are complete down to 150--250~mJy and below. This
should provide us with a sample of the same completeness
characteristics. However, it is well known that flat spectrum sources
are variable \citep[e.g.,][]{KelPau_ARAA68}; consequently, the
variability corrupts at some level our estimations of spectral
index and total flux density. The membership of a source in the
completeness sample is also changeable and depends on the observation
epochs of the various compilation surveys. The quantitative analysis of
completeness of the resulting  correlated flux density limited sample of
the sources from the combined ICRF-Ext.2 and VCS1 to VCS5 catalogs will have
to take into account the frequency dependent variability properties
\citep[e.g.,][]{KKNB02} as well as the compactness characteristics of flat
spectrum sources \citep[e.g.,][]{PopovKovalev99,2cmPaperIV}. This is
beyond the scope of the present paper and is deferred to another
publication. We expect the present sample to be sufficiently complete,
robust, and unbiased for most statistical studies of flat-spectrum radio
sources.

\section{Observations and data processing}
\label{s:obs_anal}

  The VCS5 observations were carried out in three 24-hour observing
sessions with the VLBA on 2005 July 8, July 9, and July 20. Each of the
675 target sources was observed in two scans of 120 seconds each. The
target sources were observed in a sequence designed to minimize loss of
time from antenna slewing. In addition to these objects, 97 strong
sources were taken from the GSFC astrometric catalog
\mbox{\tt{}2004f\_astro}\footnote{\url{http://vlbi.gsfc.nasa.gov/solutions/astro}}.
Observations of three or four strong sources from this list were made every
1\,h to 1.5\,h, 70\,s to 80\,s seconds per scan. These observations were scheduled
in such a way that at each VLBA station at least one of these sources
was observed at an elevation angle less than 20\degr, and at least one
at an elevation angle greater than 50\degr. The purpose of these
observations was to provide calibration for mis-modeled atmospheric path
delays and to tie the VCS5 source positions to the ICRF catalog
\citep{icrf98}. The list of tropospheric
calibrators\footnote{\url{http://vlbi.gsfc.nasa.gov/vcs/tropo\_cal.html}}
was selected from the sources that, according to the 2~cm VLBA survey
results \citep{2cmPaperIV}, showed the greatest compactness index, i.e.\
the ratio of the correlated flux density measured at long VLBA spacings
to the flux density integrated over the VLBA image. In total, 772
targets and calibrators were observed. The antennas were on-source about
65\,\% of the time.

  Similar to the previous VLBA Calibrator Survey observing campaigns
\citep[e.g.,][]{VCS4}, we used the VLBA dual-frequency geodetic mode,
observing simultaneously at S~band and X~band. Each band was separated into
four 8~MHz channels (IFs) which spanned 140 MHz around 2.3~GHz and 490
MHz around 8.6~GHz to provide precise measurements of group
delays for astrometric processing. Since the a~priori coordinates of
candidates were expected to have errors of up to 30\arcsec, the data
were correlated with an accumulation period of 1~second in 64~spectral
channels in order to provide an extra-wide window for fringe search.

  Processing of the VLBA correlator output was done in three steps. In
the first step the data were calibrated using the Astronomical Image
Processing System (AIPS) \citep{aips}. In the second step data were
imported to the Caltech DIFMAP package~\citep{difmap}, $uv$ data
flagged, and maps were produced using an automated procedure of hybrid
imaging developed by Greg Taylor~\citep{difmap-script} which we adopted
for our needs. We were able to reach the VLBA image thermal noise level
for most of our CLEAN images \citep{VLBA_summ}. Errors on our estimates
of correlated flux density values for sources stronger than
$\sim$100~mJy were dominated by the accuracy of amplitude
calibration, which for the VLBA, according to \citet{VLBA_summ}, is at
the level of 5\,\% at 1~GHz to 10~GHz. An additional error is introduced by
the fact that our frequency channels are widely spread over receiver
bands while the VLBA S~band and X~band gain-curve parameters are measured
around 2275~GHz and 8425~MHz respectively \citep{VLBA_summ}, and the noise
diode spectrum is not ideally flat. However, this should not add more
than a few percent to the total resulting error. Our error estimate was
confirmed by comparison of the flux densities integrated over the VLBA
images with the single-dish flux densities which we measured with
\mbox{RATAN--600} in June and August 2005 for slowly varying sources
without extended structure. The methods of single-dish observations and
data processing can be found in \citet{Kovalev_etal99}. In the third
step, the data were imported to the Calc/Solve program, group delays
ambiguities were resolved, outliers eliminated, and coordinates of new
sources were adjusted using ionosphere-free combinations of X~band and
S~band group delay observables of the three VCS5 sessions, 19 VCS1 to
VCS4 experiments and 3976 24-hour International VLBI Service for
astrometry and geodesy (IVS)
experiments\footnote{\url{http://vlbi.gsfc.nasa.gov/solutions/2005c}} in
a single least square solution. Positions of 3486 sources were estimated
including all detected VCS5 sources: 590 targets and 97 tropospheric
calibrators. Boundary conditions were imposed requiring zero
net-rotation of position adjustments of the 212 sources listed as
defining sources in the ICRF catalog with respect to their coordinates
from that catalog.

\begin{figure}[b]
\begin{center}
\resizebox{\hsize}{!}{
\includegraphics[trim=0cm 0cm 0cm 0cm]{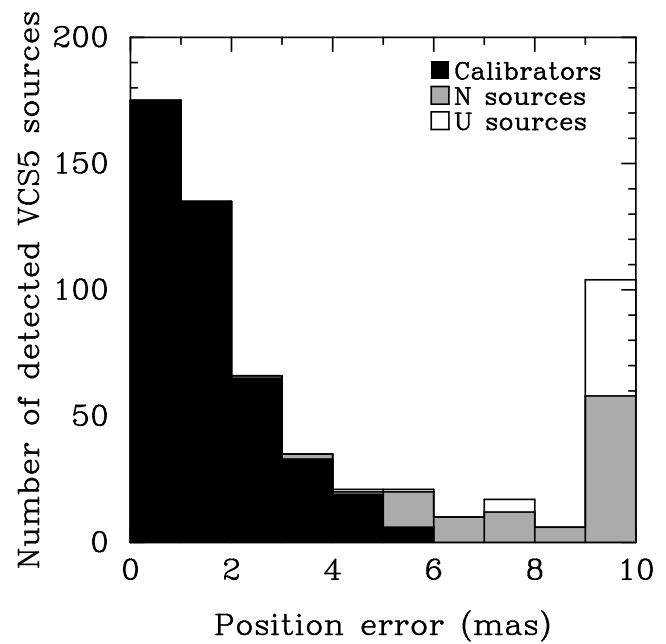}
}
\end{center}
\caption{Histogram of semi-major error ellipse of position errors.
The last bin shows errors exceeding 9~mas.
See explanation of different assigned source classes in
\S\,\ref{s:obs_anal}, \S\,\ref{s:catalog}.
\label{f:errhist}
}
\end{figure}

  In a separate solution, coordinates of the 97 well known tropospheric
calibrators were estimated from the VCS5 observing sessions only.
Comparison of these estimates with coordinates derived from the 3976 IVS
geodetic/astrometric sessions provided us a measure of the accuracy of
the coordinates from the VCS5  observing campaign. The differences in
coordinate estimates were used for computation of parameters $a$ and
$b(\delta)$ of an error inflation model in the form $ \sqrt{ (a\,
\sigma)^2 + b(\delta)^2}$, where $\sigma$ is an uncertainty derived from
the fringe amplitude signal to noise ratio using the error propagation
law and $\delta$ is declination. More details about the analysis and
imaging procedures can be found in \cite{VCS1} and \cite{VCS3}. The
histogram of source position errors is presented in
Figure~\ref{f:errhist}.

\begin{figure*}[p]
\begin{center}
\resizebox{1.0\hsize}{!}
{
  \includegraphics[trim=-0.5cm 5cm 0cm 0cm]{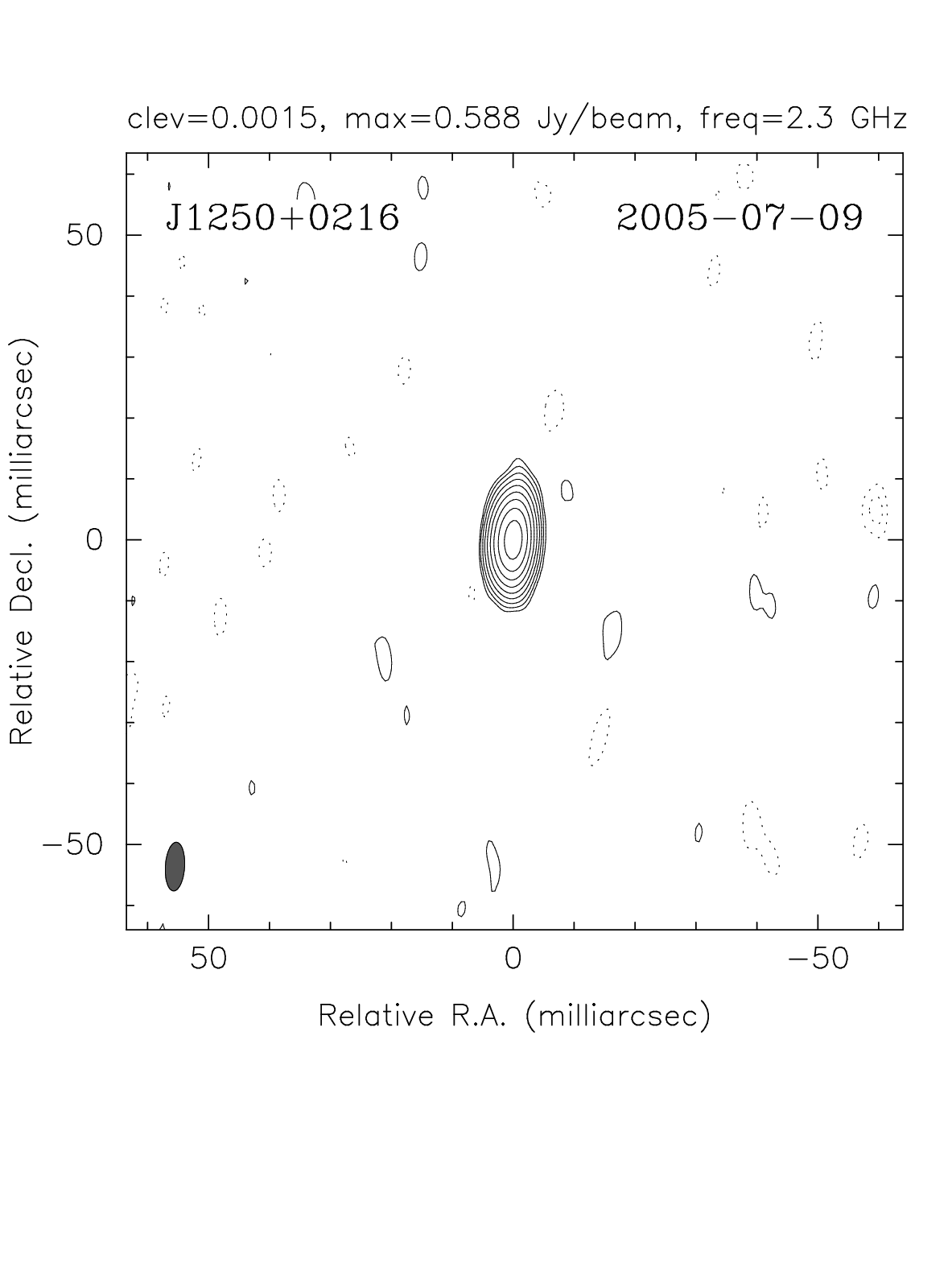}
  \includegraphics[trim=-1.0cm 5cm 0cm 0cm]{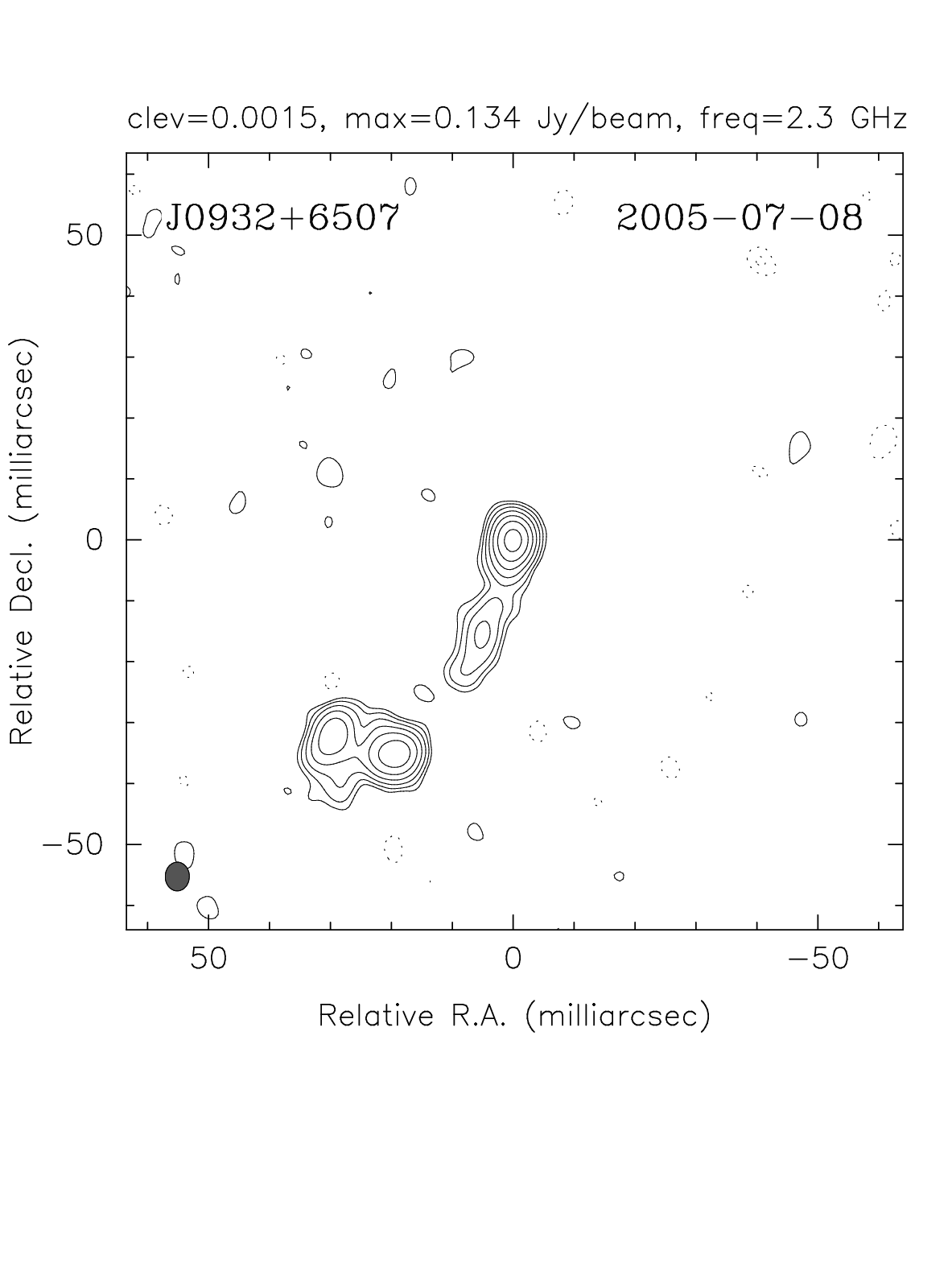}
  \includegraphics[trim=-1.0cm 5cm 0cm 0cm]{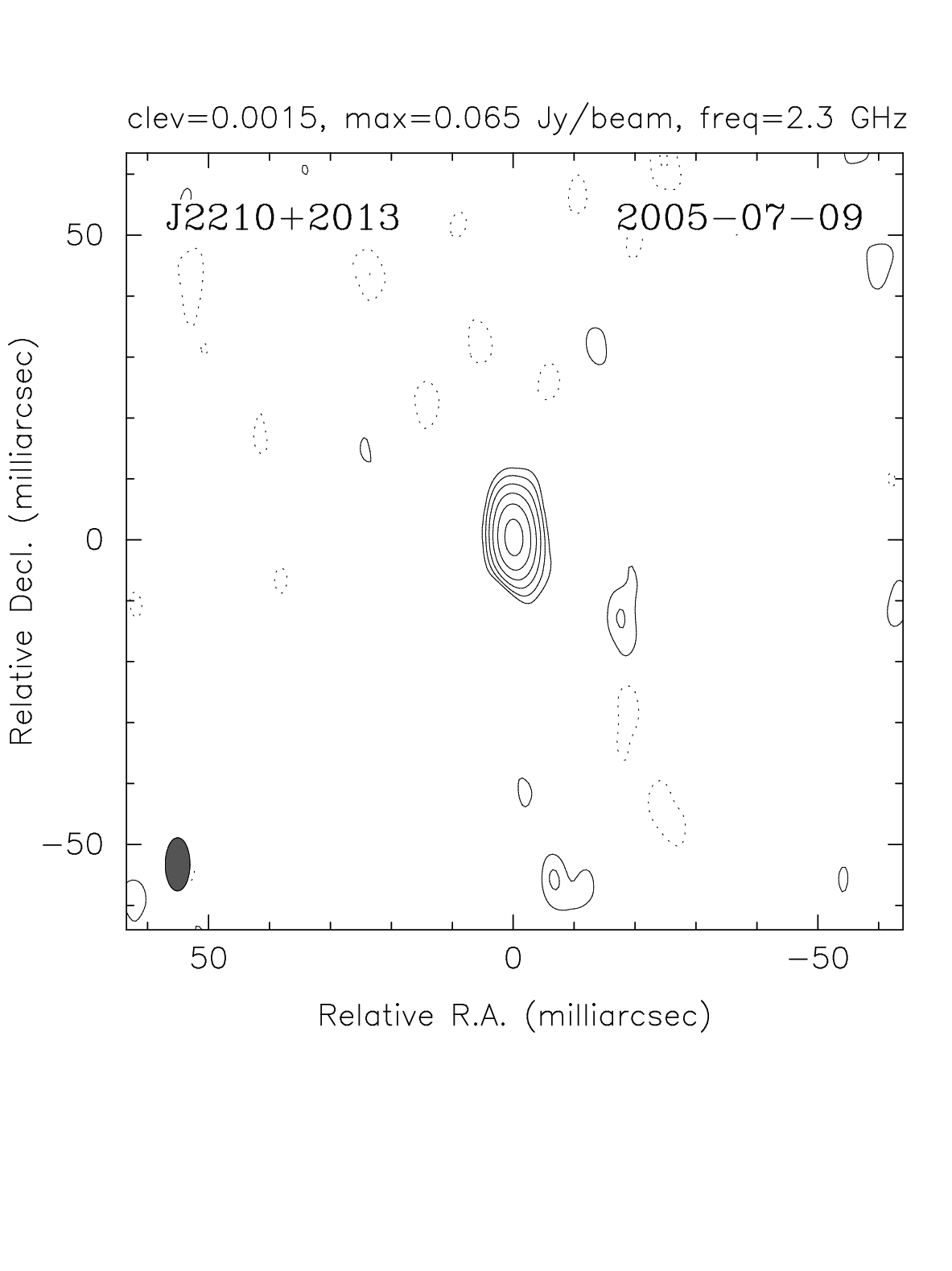}
}

\resizebox{1.0\hsize}{!}
{
  \includegraphics[trim=0cm  0.0cm 0cm 0cm,angle=270]{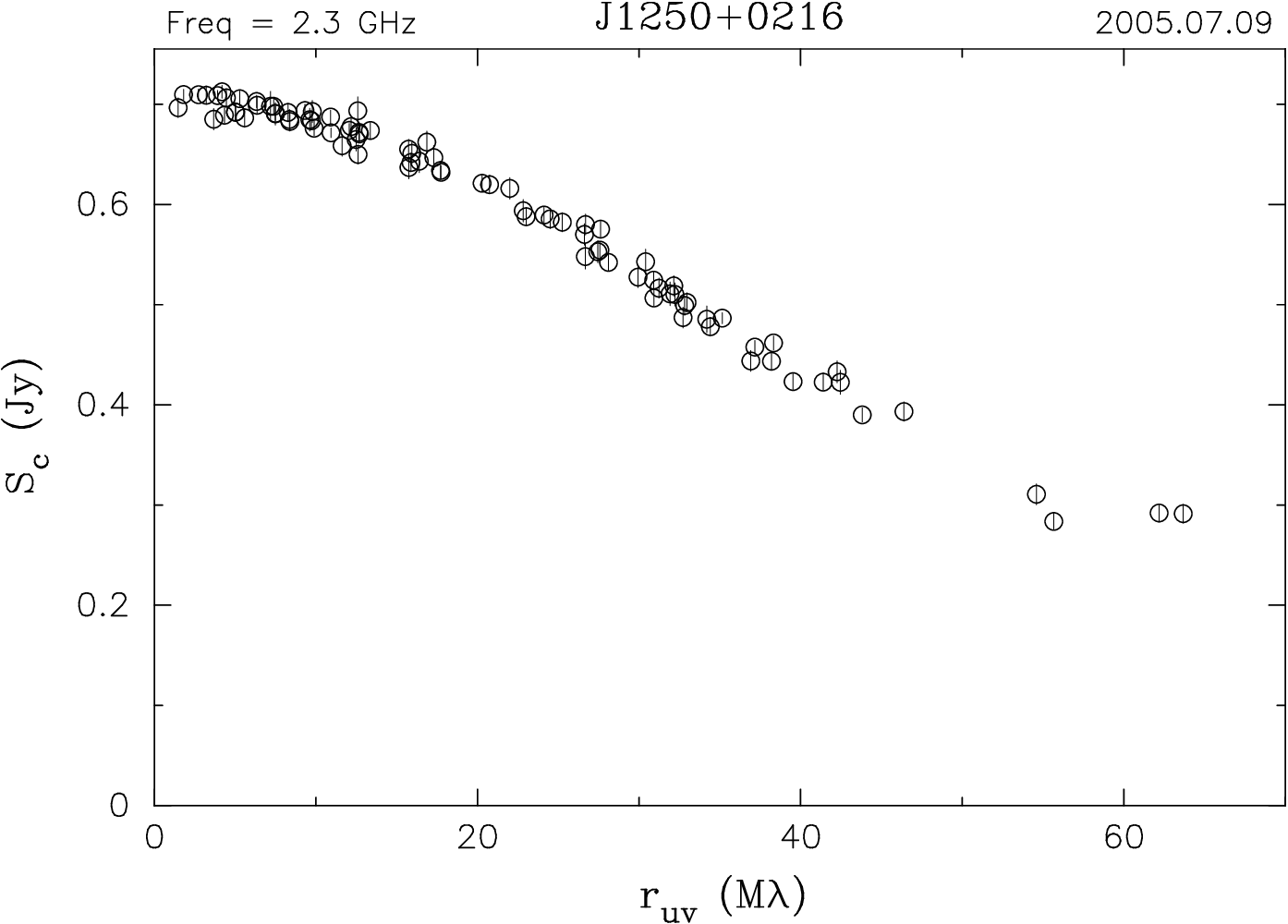}
  \includegraphics[trim=0cm -0.5cm 0cm 0cm,angle=270]{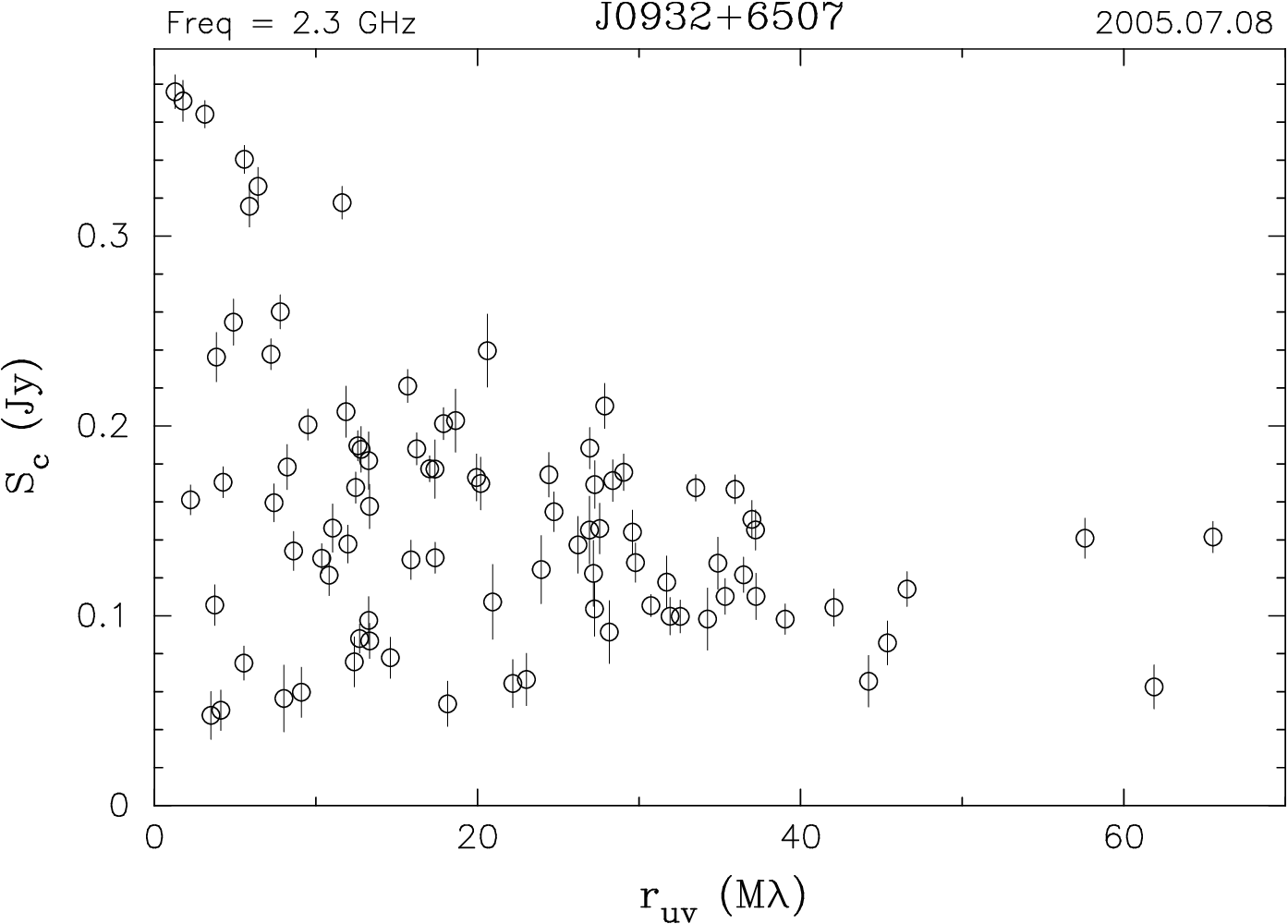}
  \includegraphics[trim=0cm -0.5cm 0cm 0cm,angle=270]{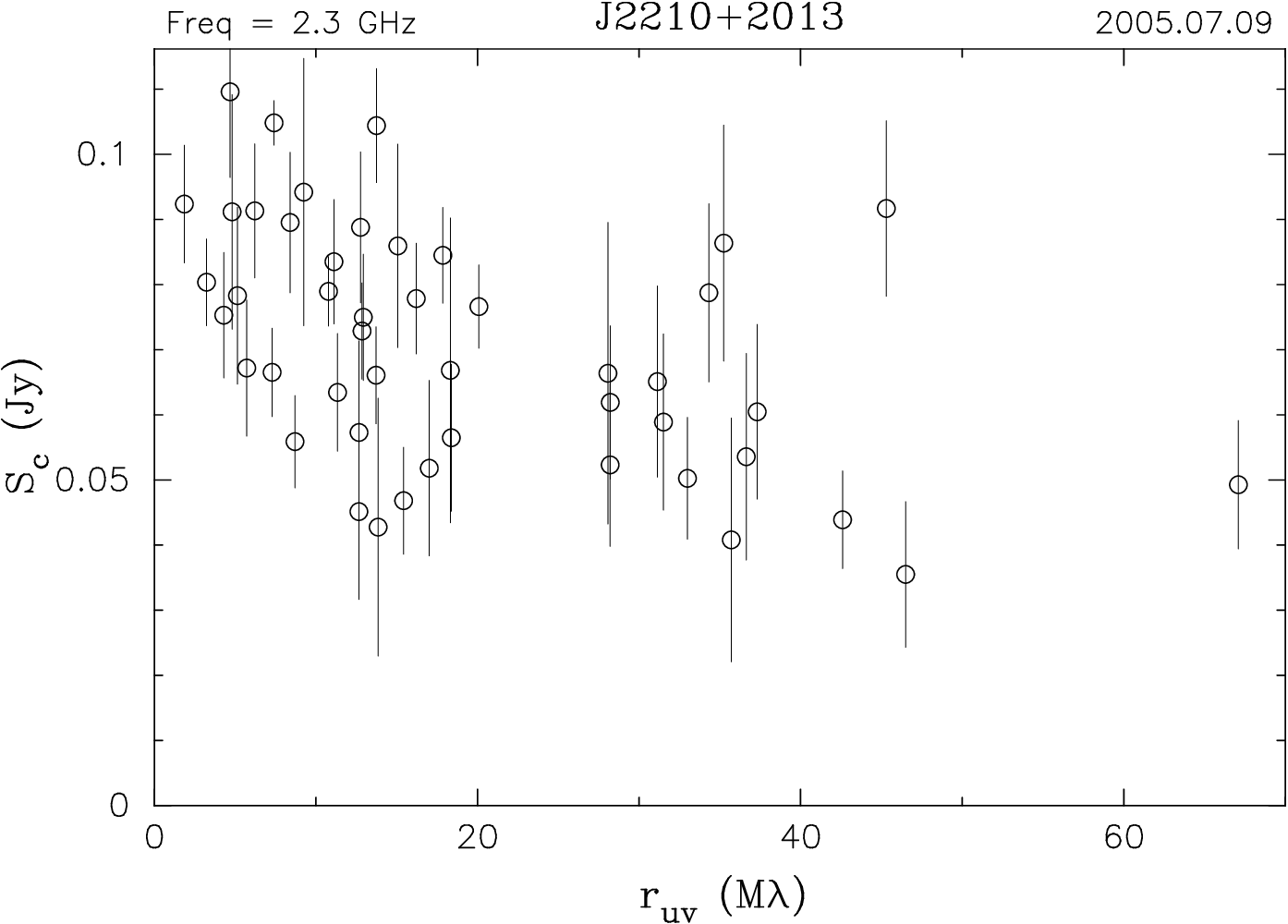}
}

\resizebox{1.0\hsize}{!}
{
  \includegraphics[trim=-0.5cm 5cm 0cm 0cm]{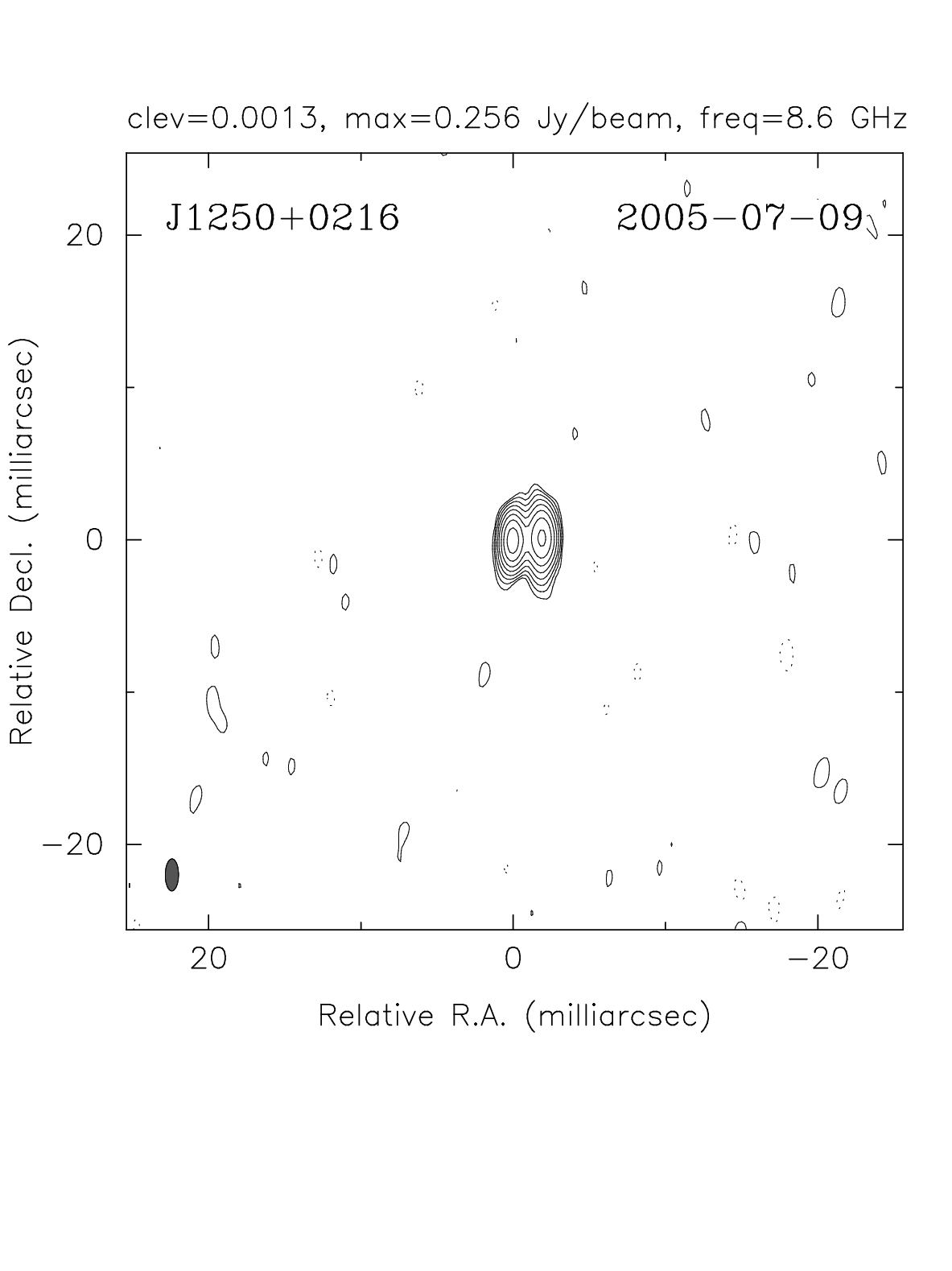}
  \includegraphics[trim=-1.0cm 5cm 0cm 0cm]{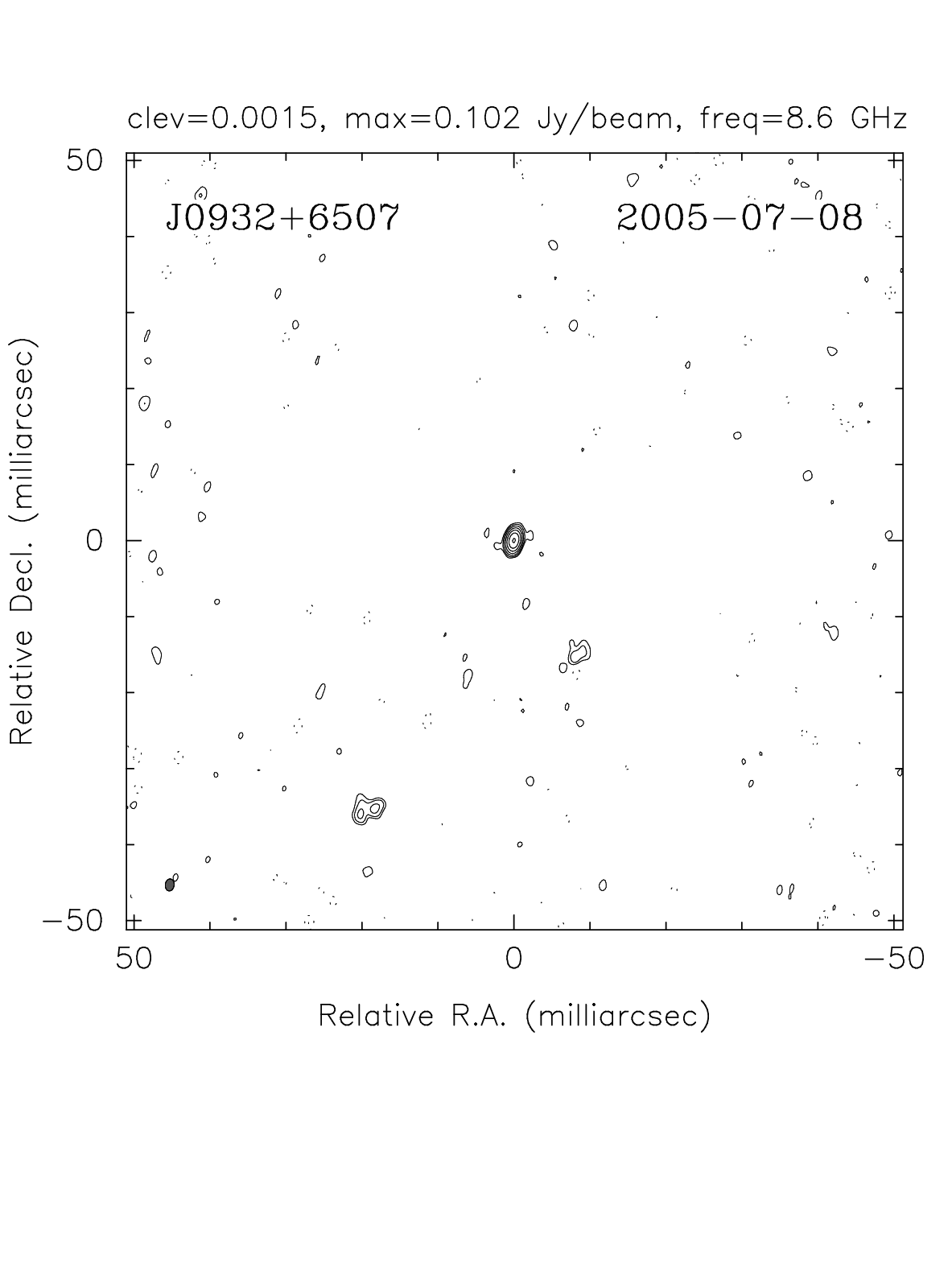}
  \includegraphics[trim=-1.0cm 5cm 0cm 0cm]{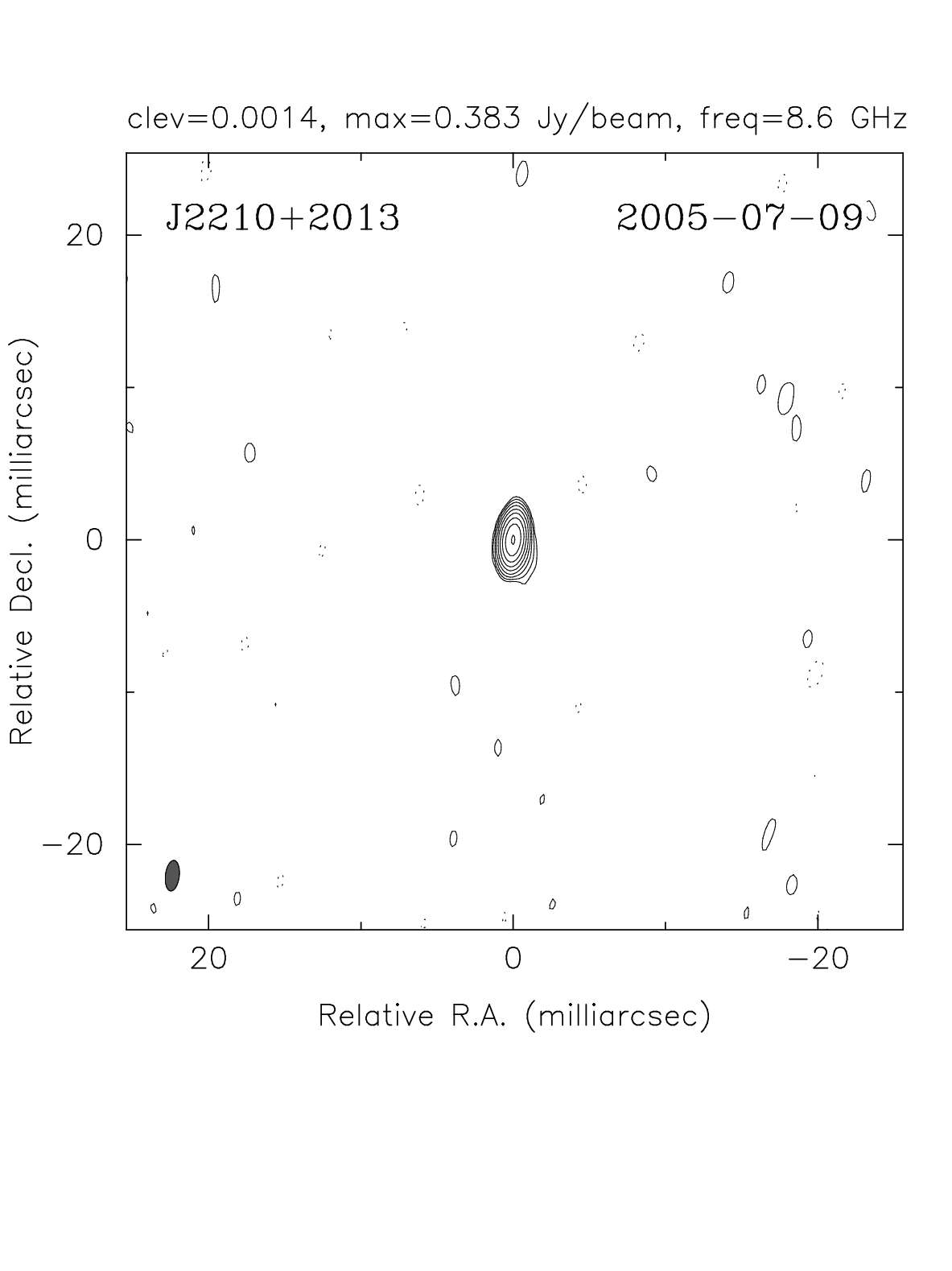}
}

\resizebox{1.0\hsize}{!}
{
  \includegraphics[trim=0cm  0.0cm 0cm 0cm,angle=270]{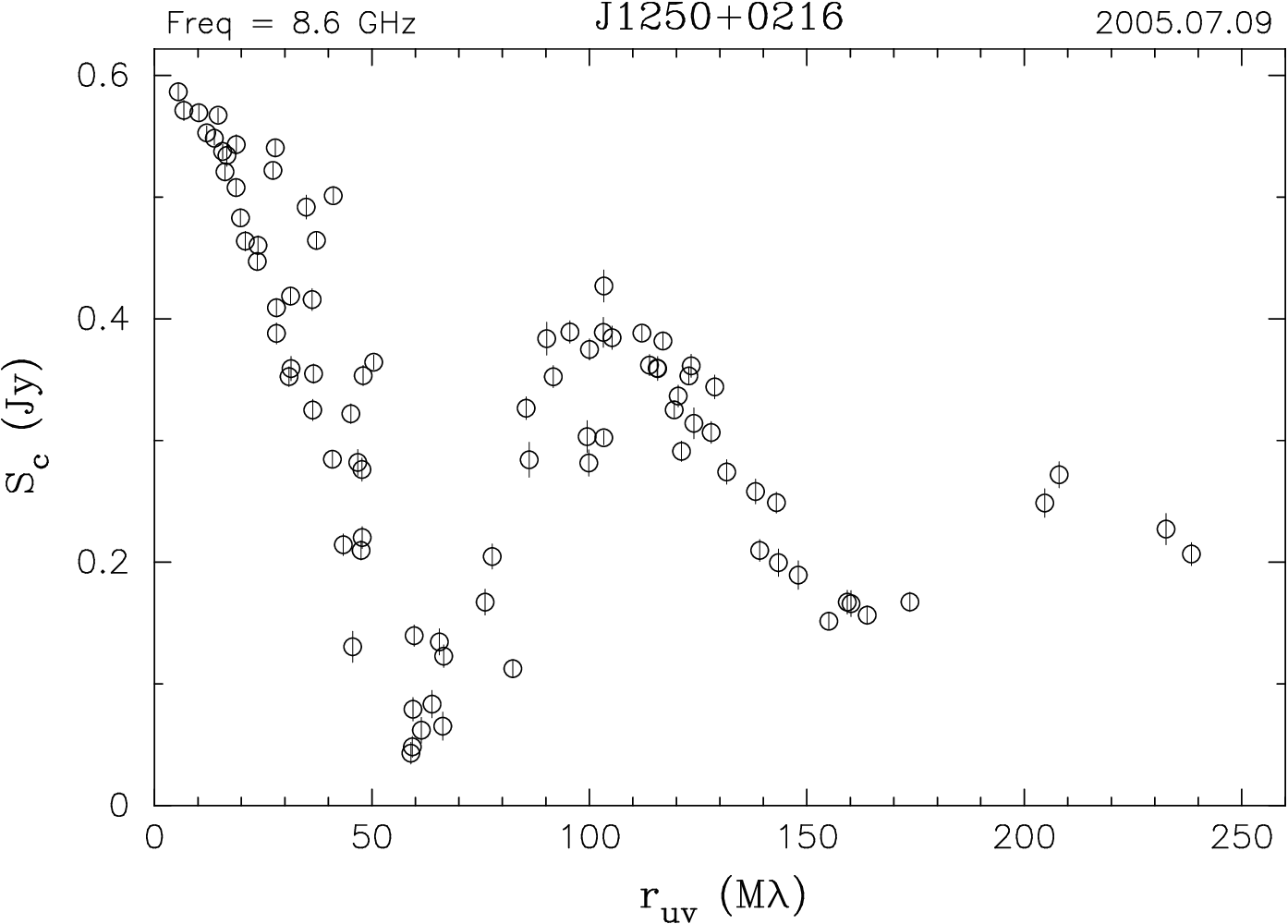}
  \includegraphics[trim=0cm -0.5cm 0cm 0cm,angle=270]{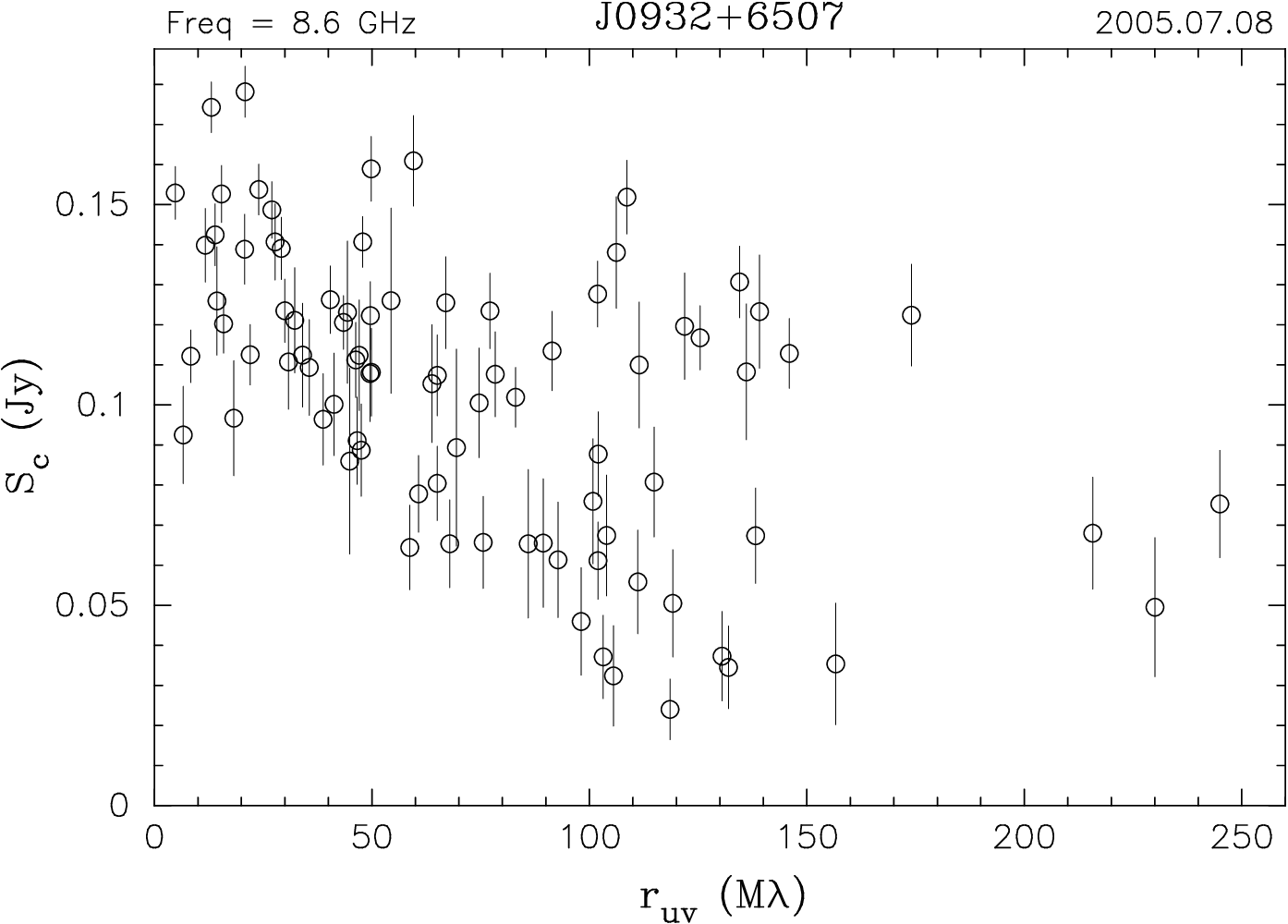}
  \includegraphics[trim=0cm -0.5cm 0cm 0cm,angle=270]{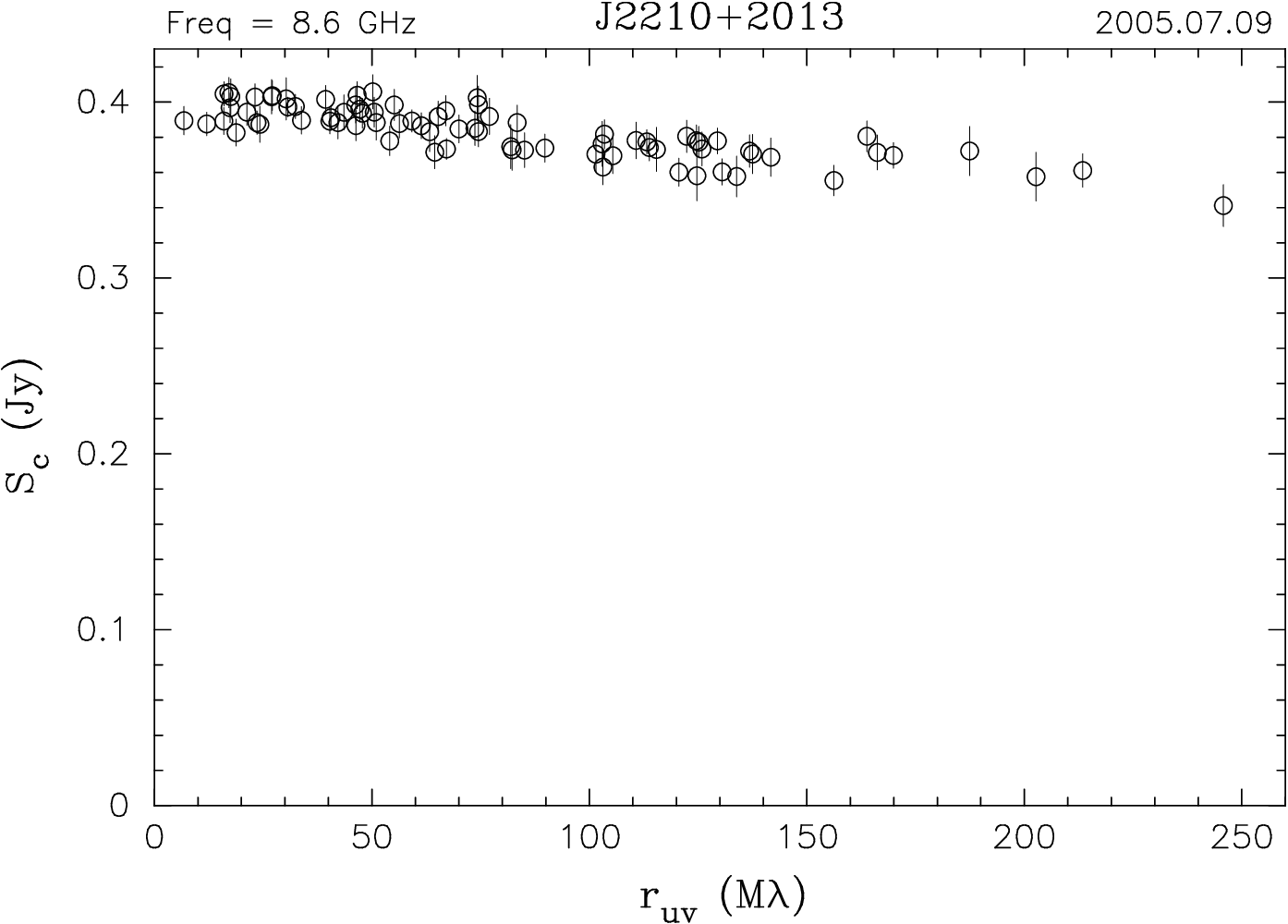}
}
\end{center}
\figcaption{\footnotesize 
From top to bottom.
{\em Row~1:}
Naturally weighted CLEAN images at S~band (2.3~GHz). The lowest contour
is plotted at the leven given by ``clev'' in each panel title (Jy/beam),
the peak brightness is given by ``max'' (Jy/beam). The contour levels
increase by factors of two. The dashed contours indicate negative flux.
The beam is shown in the bottom left corner of the images.
{\em Row~2:}
Dependence of the correlated flux density at S~band on projected
spacing. Each point represents a coherent
average over one 2~min observation on an individual interferometer
baseline. The error bars represent only the statistical errors.
{\em Row~3:} Naturally
weighted CLEAN images at X~band (8.6~GHz).
{\em Row~4:} Dependence of the
correlated flux density at X~band on projected spacing.
\label{f:images}
}
\end{figure*}

  In total, 590 out of 675 sources were detected and yielded at least
two good points for position determination. This 87\,\% detection rate 
confirms the validity of the applied candidate selection procedure
(\S\ref{s:selection}). It should be noted that, due to an omission, the
list of target sources contained 21 objects previously observed and
detected in the VCS4 campaign.

However, not all of these 590 sources are suitable as phase reference
calibrators or as targets for geodetic observations. Following
\citet{VCS3} we consider a source suitable as a calibrator if 1) the
number of good X/S pairs of observations is eight or greater in order to
rule out the possibility of a group delay ambiguity resolution error,
and 2) the position error before re-weighting is less than 5~mas
following the strategy adopted in processing VCS observations. Only 433
sources satisfy this calibrator criteria. Other detected sources were
somewhat resolved and/or below the detection limit of these observations
of 60 mJy. Some of these may become suitable phase calibrators for
future experiments with higher data rates and more sensitivity than the
VCS surveys. Among the 157 non-calibrators, 53 sources  had less than
eight observations at X~band and less than eight observations S~band.
Positions  of these sources we consider as unreliable, because we 
cannot rule out errors in group delay ambiguity resolution. The other 104
sources had eight or more observations at one of the bands, so we can rule
out the possibility of group delay ambiguity  resolution errors, and
therefore, we consider that our estimates of positions of these sources
are reliable.

\section{The VCS5 catalog}
\label{s:catalog}

\begin{figure}[t]
\begin{center}
\resizebox{1.0\hsize}{!}
{
  \includegraphics[trim= 0.00cm 0cm 0cm 0cm]{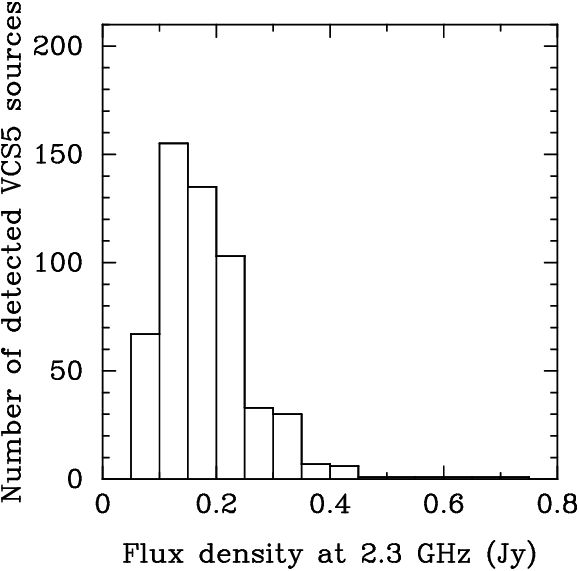}
  \includegraphics[trim= 1.5cm 0cm 0cm 0cm,clip]{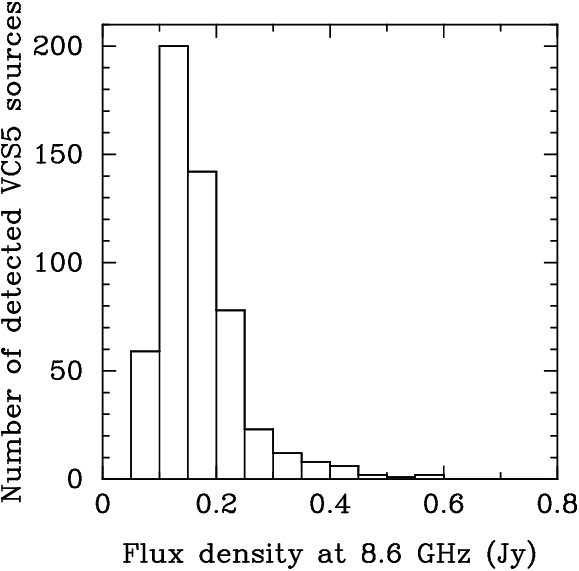}
}
\end{center}
\figcaption{
Distributions of flux density integrated over VLBA image for all
detected VCS5 sources (columns 10 and 12 of the Table~\ref{t:cat}).
\label{f:hist_s_spi}
}
\end{figure}

The VCS5 catalog of 590 detected target sources is listed in
Table~\ref{t:cat}. The first column gives source class: ``C'' if the
source can be used as a calibrator, ``N'' if it cannot but determined
positions are reliable, ``U''---non-calibrator, unreliable positions.
The second and third columns give IVS source name (B1950 notation), and
IAU name (J2000 notation). The fourth and fifth columns give measured
source coordinates at the J2000.0 epoch. Columns 6 and 7 give inflated
source position uncertainties in right ascension (without $\cos\delta$
factor) and declination in mas, and column 8 gives the correlation
coefficient between the errors in right ascension and declination. The
number of group delays used for position determination is listed in
column 9. Columns 10 and 12 give the estimate of the flux density
integrated over the entire map in janskies at X~band and S~band
respectively. This estimate was computed as a sum of all CLEAN
components if a CLEAN image was produced. If we did not have enough
detections of the visibility function to produce a reliable image, the
integrated flux density was estimated as the median of the correlated
flux density measured at projected spacings less than 25~M$\lambda$ and
7~M$\lambda$ for X~band and S~bands respectively.  The integrated flux
density means the total flux density with spatial frequencies less than
4~M$\lambda$ at X~band and 1~M$\lambda$ at S~band filtered out, or in
other words, the flux density from all components within a region about
or less than 50~mas at X~band and 200~mas at S~band. Columns 11 and 13
give the flux density of unresolved components estimated as the median
of correlated flux density values measured at projected spacings greater
than 170~M$\lambda$ for X~band and greater than 45~M$\lambda$ for
S~band. For some sources no estimates of the integrated and/or
unresolved flux density are presented, because either no data were
collected on the baselines used in the calculations, or these data were
unreliable. Column 14 gives the data type used for position estimation:
X/S stands for ionosphere-free linear combination of X and S wide-band
group delays; X stands for X-band-only group delays; and S stands for
S-band-only group delays. Some sources for which less than eight pairs
of X~band and S~band group delay observables were available had two or
more observations at X~band and/or S~band. For these sources either
X~band or S~band only estimates of coordinates are listed in the VCS5
catalog.

\begin{figure}[t]
\begin{center}
\resizebox{1.0\hsize}{!}
{
  \includegraphics[trim=0cm 0cm 0cm 0cm,angle=270]{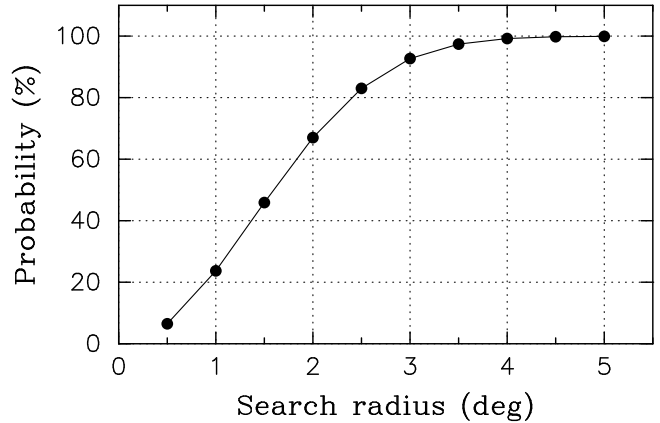}
}
\end{center}
\figcaption{
The probability (filled circles) of finding a calibrator in any given
direction within a circle of a given radius, north of declination
$-30\degr$. All sources from 3976 IVS geodetic/astrometric sessions and
22 VCS1 to VCS5 VLBA sessions that are classified as calibrators are taken
into account.
\label{f:prob}
}
\end{figure}

  In addition to this table, the html version of the catalog is posted
on the Web\footnote{\url{http://vlbi.gsfc.nasa.gov/vcs5}}. For each
source there are eight links: to a pair of postscript images of the
source at X~band and S~band; a pair of plots of correlated flux
density as a function of baseline length projected to the source plane;
a pair of FITS files of CLEAN components of naturally weighted source
images; and to a pair of FITS files with calibrated $uv$ data. This
dataset is also accessible from the NRAO
archive\footnote{\url{http://archive.nrao.edu}} which links the files to
the Virtual Observatory. The positions and the plots are also accessible
from the updated NRAO VLBA Calibrator Search
web-page\footnote{\url{http://www.vlba.nrao.edu/astro/calib}}.

\begin{deluxetable*}{c l l l r r r r r r r r r c}
\tablewidth{0pt}
\tablecaption{\rm The VCS5 catalog \label{t:cat}}
\tabletypesize{\scriptsize}
\tablehead{
   \colhead{} &
   \multicolumn{2}{c}{Source name}                   &
   \multicolumn{2}{c}{J2000.0 Coordinates}           &
   \multicolumn{2}{c}{Errors (mas)}                  &
   \colhead{}                                        &
   \colhead{}                                        &
   \multicolumn{4}{c}{Correlated flux density (Jy)}  &
   \colhead{}
   \vspace{0.5ex} \\
   \multicolumn{9}{c}{}         &
   \multicolumn{2}{c}{8.6 GHz}  &
   \multicolumn{2}{c}{2.3 GHz}  \vspace{0.5ex} \\
   \colhead{Class}    &
   \colhead{IVS}      &
   \colhead{IAU}      &
   \colhead{Right ascension} &
   \colhead{Declination}     &
   \colhead{$\Delta \alpha$} &
   \colhead{$\Delta \delta$} &
   \colhead{Corr}   &
   \colhead{\# Obs} &
   \colhead{Total } &
   \colhead{Unres } &
   \colhead{Total } &
   \colhead{Unres } &
   \colhead{Band}   
   \vspace{0.5ex} \\
   \colhead{(1)}    &
   \colhead{(2)}    &
   \colhead{(3)}    &
   \colhead{(4)}    &
   \colhead{(5)}    &
   \colhead{(6)}    &
   \colhead{(7)}    &
   \colhead{(8)}    &
   \colhead{(9)}    &
   \colhead{(10)}   &
   \colhead{(11)}   &
   \colhead{(12)}   &
   \colhead{(13)}   &
   \colhead{(14)} 
   }
\startdata
C  & \objectname{0008$+$006} & \objectname{J0011$+$0057}  & 00 11 30.403309 & $+$00 57 51.87984  &   1.02 &   2.00 &     0.114 &     25 &   0.09 &    0.07 &      \n &      \n &  X   \\
C  & \objectname{0009$+$467} & \objectname{J0012$+$4704}  & 00 12 29.302900 & $+$47 04 34.73946  &   0.77 &   1.08 &  $-$0.371 &     35 &   0.13 &    0.12 &    0.10 & $<$0.06 &  X/S \\
N  & \objectname{0013$-$240} & \objectname{J0016$-$2343}  & 00 16 05.738818 & $-$23 43 52.18956  &  30.58 &  16.85 &  $-$0.808 &     17 &     \n &      \n &    0.15 &    0.08 &  S   \\
C  & \objectname{0015$-$054} & \objectname{J0017$-$0512}  & 00 17 35.817204 & $-$05 12 41.76727  &   0.46 &   0.92 &  $-$0.278 &     54 &   0.20 &    0.12 &    0.14 &    0.09 &  X/S \\
C  & \objectname{0015$-$280} & \objectname{J0017$-$2748}  & 00 17 59.006128 & $-$27 48 21.57153  &   1.75 &   3.51 &     0.712 &     32 &   0.24 &    0.16 &    0.20 &    0.06 &  X/S \\
C  & \objectname{0034$+$078} & \objectname{J0037$+$0808}  & 00 37 32.197173 & $+$08 08 13.05750  &   0.38 &   0.50 &  $-$0.225 &     76 &   0.25 &    0.14 &    0.19 &    0.10 &  X/S \\
C  & \objectname{0035$-$037} & \objectname{J0038$-$0329}  & 00 38 20.794340 & $-$03 29 58.96178  &   0.32 &   0.63 &  $-$0.347 &     77 &   0.20 &    0.16 &    0.31 &    0.21 &  X/S \\
N  & \objectname{0036$-$191} & \objectname{J0039$-$1854}  & 00 39 16.924431 & $-$18 54 05.61863  &   4.97 &   9.83 &     0.749 &      8 &   0.12 &      \n &    0.18 &    0.10 &  X/S \\
C  & \objectname{0037$+$011} & \objectname{J0040$+$0125}  & 00 40 13.525489 & $+$01 25 46.35014  &   0.99 &   1.94 &  $-$0.124 &     29 &   0.11 & $<$0.06 & $<$0.06 &      \n &  X   \\
C  & \objectname{0041$+$677} & \objectname{J0044$+$6803}  & 00 44 50.759589 & $+$68 03 02.68607  &   0.91 &   0.58 &  $-$0.130 &     59 &   0.28 &    0.22 &    0.14 & $<$0.06 &  X/S \\
\enddata
\tablecomments{
Table~\ref{t:cat} is presented in its entirety in the electronic edition
of the Astronomical Journal. A portion is shown here  for guidance
regarding its form and contents. Assigned source class in (1) is `C' for
calibrator, `N' for non-calibrator with reliable coordinates,
`U' for non-calibrator with unreliable coordinates; see
\S\,\ref{s:obs_anal}, \S\,\ref{s:catalog} for details. Units of right ascension
are hours, minutes and seconds; units of declination are degrees,
minutes and seconds.
}
\end{deluxetable*}

\begin{deluxetable}{l l c c}
\tablewidth{0pt}
\tablecaption{\rm Sources not detected in VCS5 VLBA observations\label{t:nondetected}}
\tabletypesize{\scriptsize}
\tablehead{
   \multicolumn{2}{c}{Source name}                   &
   \multicolumn{2}{c}{J2000.0 Coordinates}           
   \vspace{0.5ex} \\
   \colhead{IVS}      &
   \colhead{IAU}      &
   \colhead{Right ascension} &
   \colhead{Declination} 
   \vspace{0.5ex} \\
   \colhead{(1)}    &
   \colhead{(2)}    &
   \colhead{(3)}    &
   \colhead{(4)}     
   }
\startdata
\objectname{0032$-$011} & \objectname{J0034$-$0054} & 00 34 43.93 & $-$00 54 13.0 \\
\objectname{0039$+$211} & \objectname{J0041$+$2123} & 00 41 45.10 & $+$21 23 41.1 \\
\objectname{0104$+$650} & \objectname{J0107$+$6521} & 01 07 51.35 & $+$65 21 20.5 \\
\objectname{0137$-$273} & \objectname{J0139$-$2705} & 01 39 55.42 & $-$27 05 29.4 \\
\objectname{0212$-$214} & \objectname{J0214$-$2113} & 02 14 40.73 & $-$21 13 28.2 \\
\objectname{0258$+$356} & \objectname{J0301$+$3551} & 03 01 47.96 & $+$35 51 24.5 \\
\objectname{0426$+$351} & \objectname{J0430$+$3516} & 04 30 14.41 & $+$35 16 23.1 \\
\objectname{0434$-$225} & \objectname{J0436$-$2226} & 04 36 34.31 & $-$22 26 18.6 \\
\objectname{0506$-$196} & \objectname{J0508$-$1935} & 05 08 19.03 & $-$19 35 56.4 \\
\objectname{0512$-$129} & \objectname{J0515$-$1255} & 05 15 17.51 & $-$12 55 27.8 \\
\enddata
\tablecomments{
Table~\ref{t:nondetected} is presented in its entirety in the electronic
edition of the Astronomical Journal. A portion is shown here  for
guidance regarding its form and contents. Units of right ascension are
hours, minutes and seconds; units of declination are degrees, minutes
and seconds. The 85 sources presented in this electronic table include
the gravitation lens \objectname{1422+231} which was detected, but not
processed in VCS5. The J2000 source positions are taken from the NVSS
survey \citep{NVSS}, they were used for VCS5 VLBA observing and
correlation.
}
\end{deluxetable}

Table~\ref{t:nondetected} presents 85 sources not detected in VCS5 VLBA
observations. Source positions used for observations and correlation are
presented. They are taken from \cite{NVSS}. The correlated flux density
for the non-detected sources is estimated to be less than 60~mJy at 2.3~GHz
and 8.6~GHz.

Figure~\ref{f:images} presents examples of naturally weighted contour
CLEAN images as well as correlated flux density versus projected spacing
dependence for three sources: the strongest VCS5 source at X~band,
\objectname{J1250$+$0216}, with two bright components resolved at X~band and
not resolved at S~band; a steep spectrum source with extended structure,
\objectname{J0932$+$6507}; and the source with the most inverted
spectrum and very compact structure at the milliarcsecond scale,
\objectname{J2210$+$2013}.

Figure~\ref{f:hist_s_spi} presents histograms of the 2.3~GHz and 8.6~GHz
integrated flux density for 590 detected VCS5 sources, 132 out of which
have integrated flux density $S\ge 200$~mJy at 8.6~GHz. Their addition
to the previously observed sources will form the statistically complete
sample north of declination $-30\degr$. It is interesting to note that
many of the discovered VCS5 sources have inverted radio spectra. The
50~mJy cutoff for the original selection of sources from the NVSS
catalog allowed us to add inverted-spectrum objects to the list of
candidates. A few tens of new compact VCS5 objects with high flux
density on VLBA baselines will be useful for geodetic applications. 

   The sky calibrator density for different radii of a search circle for
declination $\delta>-30\degr$ is presented in Figure~\ref{f:prob}. As
discussed in \cite{VCS4}, the addition of these sources to the VLBA
Calibrator list did not affect significantly the density for the search
radius of 4\degr, but increases it for smaller search circles, e.g.,
the probability of finding a calibrator within $2\fdg5$ is now 83\,\%. This
is beneficial for many applications requiring bright compact calibrator
within 2\degr to 3\degr of a target.

\section{Summary}
\label{s:summary}

The VCS5 Survey has made a significant step towards constructing a
homogeneous statistically complete sample of flat-spectrum compact
extragalactic radio sources north of declination $-30\degr$ with
integrated VLBA flux density greater than about 200~mJy at 8~GHz. The
VCS5 Survey has added 569 new sources, not previously observed with VLBI
under geodesy and absolute astrometry programs. Among them, 433 sources
are suitable as phase referencing calibrators and as target sources for
geodetic applications. After processing the VCS5 experiments, the total
number of sources with positions known at the  nanoradian level is 3486,
and the number of VLBI calibrators has  grown from 2472 to 2905.  This
pool of calibrators was formed from analysis of 22 VLBA Calibrator
Survey and 3976 24-hour IVS astrometry and geodesy observing
sessions.

In the present paper we do not yet provide quantitative estimates of
completeness of our list of compact flat-spectrum sources. In order to
get these estimates we are going to (i) complete the homogeneous imaging
of all of 3486 sources and get estimates of their integrated flux
densities at milliarcsecond scales in the X~band and S~band, (ii) complete
processing of instantaneous single-dish multi-frequency, multi-epoch flux
density measurements with \mbox{RATAN--600} for this sample, (iii)
observe with the VLBA a total-flux-density limited sample of all sources
regardless of their spectral index over a relatively large portion of
the sky complemented with simultaneous multi-frequency single-dish
measurements. The latter will allow us to assess whether conclusions
drawn from the VLBI flat-spectrum source samples can be extended to the
whole population of extragalactic objects regardless their continuum
spectrum characteristics.



\acknowledgments

   We would like to thank A.~Roy and the anonymous referee for useful
comments which helped to improve the manuscript.
     \facility[NRAO(VLBA)]{The National Radio Astronomy Observatory is a
facility of the National Science Foundation operated under cooperative
agreement by Associated Universities, Inc.} We thank the staff of the
VLBA for carrying these observations in their usual efficient manner.
     Y.~Y.~Kovalev is a Research Fellow of the Alexander von Humboldt
Foundation. This work was done while D.~Gordon worked for Raytheon,
under NASA contract NAS5--01127.  \mbox{RATAN--600} observations were
partly supported by the  Russian Ministry of Education and Science, the
NASA JURRISS Program (project W--19611), and the Russian Foundation for
Basic Research (projects 01--02--16812 and 05--02--17377). The authors
made use of the database CATS \citep{CATS} of the Special Astrophysical
Observatory. This research has made use of the NASA/IPAC Extragalactic
Database (NED) which is operated by the Jet Propulsion Laboratory,
California Institute of Technology, under contract with the National
Aeronautics and Space Administration.

\bibliographystyle{apj}
\bibliography{yyk}

\begin{thebibliography}{28}
\expandafter\ifx\csname natexlab\endcsname\relax\def\natexlab#1{#1}\fi

\bibitem[{{Beasley} {et~al.}(2002){Beasley}, {Gordon}, {Peck}, {Petrov},
  {MacMillan}, {Fomalont}, \& {Ma}}]{VCS1}
{Beasley}, A.~J., {Gordon}, D., {Peck}, A.~B., {Petrov}, L., {MacMillan},
  D.~S., {Fomalont}, E.~B., \& {Ma}, C. 2002, \apjs, 141, 13

\bibitem[{{Browne} {et~al.}(1998){Browne}, {Wilkinson}, {Patnaik}, \&
  {Wrobel}}]{JVAS2}
{Browne}, I.~W.~A., {Wilkinson}, P.~N., {Patnaik}, A.~R., \& {Wrobel}, J.~M.
  1998, \mnras, 293, 257

\bibitem[{{Condon} {et~al.}(1998){Condon}, {Cotton}, {Greisen}, {Yin},
  {Perley}, {Taylor}, \& {Broderick}}]{NVSS}
{Condon}, J.~J., {Cotton}, W.~D., {Greisen}, E.~W., {Yin}, Q.~F., {Perley},
  R.~A., {Taylor}, G.~B., \& {Broderick}, J.~J. 1998, \aj, 115, 1693

\bibitem[{{Fey} {et~al.}(2004){Fey}, {Ma}, {Arias}, {Charlot},
  {Feissel-Vernier}, {Gontier}, {Jacobs}, {Li}, \&
  {MacMillan}}]{icrf-ext2-2004}
{Fey}, A.~L., {Ma}, C., {Arias}, E.~F., {Charlot}, P., {Feissel-Vernier}, M.,
  {Gontier}, A.-M., {Jacobs}, C.~S., {Li}, J., \& {MacMillan}, D.~S. 2004, \aj,
  127, 3587

\bibitem[{{Fomalont} {et~al.}(2003){Fomalont}, {Petrov}, {MacMillan}, {Gordon},
  \& {Ma}}]{VCS2}
{Fomalont}, E.~B., {Petrov}, L., {MacMillan}, D.~S., {Gordon}, D., \& {Ma}, C.
  2003, \aj, 126, 2562

\bibitem[{{Gregory} \& {Condon}(1991)}]{87GB}
{Gregory}, P.~C., \& {Condon}, J.~J. 1991, \apjs, 75, 1011

\bibitem[{{Gregory} {et~al.}(1996){Gregory}, {Scott}, {Douglas}, \&
  {Condon}}]{GB6}
{Gregory}, P.~C., {Scott}, W.~K., {Douglas}, K., \& {Condon}, J.~J. 1996,
  \apjs, 103, 427

\bibitem[{{Greisen}(2003)}]{aips}
{Greisen}, E.~W. 2003, in Astrophysics and Space Science Library 285,
  Information Handling in Astronomy -- Historical Vistas, ed. A.~{Heck}
  (Dordrecht: Kluwer), 109

\bibitem[{{Griffith} {et~al.}(1995){Griffith}, {Wright}, {Burke}, \&
  {Ekers}}]{Griffith_etal95}
{Griffith}, M.~R., {Wright}, A.~E., {Burke}, B.~F., \& {Ekers}, R.~D. 1995,
  \apjs, 97, 347

\bibitem[{{Kellermann} \& {Pauliny-Toth}(1968)}]{KelPau_ARAA68}
{Kellermann}, K.~I., \& {Pauliny-Toth}, I.~I.~K. 1968, \araa, 6, 417

\bibitem[{{Kovalev} {et~al.}(2005){Kovalev}, {Kellermann}, {Lister}, {Homan},
  {Vermeulen}, {Cohen}, {Ros}, {Kadler}, {Lobanov}, {Zensus}, {Kardashev},
  {Gurvits}, {Aller}, \& {Aller}}]{2cmPaperIV}
{Kovalev}, Y.~Y., {Kellermann}, K.~I., {Lister}, M.~L., {Homan}, D.~C.,
  {Vermeulen}, R.~C., {Cohen}, M.~H., {Ros}, E., {Kadler}, M., {Lobanov},
  A.~P., {Zensus}, J.~A., {Kardashev}, N.~S., {Gurvits}, L.~I., {Aller}, M.~F.,
  \& {Aller}, H.~D. 2005, \aj, 130, 2473

\bibitem[{{Kovalev} {et~al.}(2002){Kovalev}, {Kovalev}, {Nizhelsky}, \&
  {Bogdantsov}}]{KKNB02}
{Kovalev}, Y.~Y., {Kovalev}, Y.~A., {Nizhelsky}, N.~A., \& {Bogdantsov}, A.~B.
  2002, Publications of the Astronomical Society of Australia, 19, 83

\bibitem[{{Kovalev} {et~al.}(1999){Kovalev}, {Nizhelsky}, {Kovalev}, {Berlin},
  {Zhekanis}, {Mingaliev}, \& {Bogdantsov}}]{Kovalev_etal99}
{Kovalev}, Y.~Y., {Nizhelsky}, N.~A., {Kovalev}, Y.~A., {Berlin}, A.~B.,
  {Zhekanis}, G.~V., {Mingaliev}, M.~G., \& {Bogdantsov}, A.~V. 1999, \aaps,
  139, 545

\bibitem[{{Ma} {et~al.}(1998){Ma}, {Arias}, {Eubanks}, {Fey}, {Gontier},
  {Jacobs}, {Sovers}, {Archinal}, \& {Charlot}}]{icrf98}
{Ma}, C., {Arias}, E.~F., {Eubanks}, T.~M., {Fey}, A.~L., {Gontier}, A.-M.,
  {Jacobs}, C.~S., {Sovers}, O.~J., {Archinal}, B.~A., \& {Charlot}, P. 1998,
  \aj, 116, 516

\bibitem[{{Myers} {et~al.}(2003){Myers}, {Jackson}, {Browne}, {de Bruyn},
  {Pearson}, {Readhead}, {Wilkinson}, {Biggs}, {Blandford}, {Fassnacht},
  {Koopmans}, {Marlow}, {McKean}, {Norbury}, {Phillips}, {Rusin}, {Shepherd},
  \& {Sykes}}]{Myers_etal03}
{Myers}, S.~T., {Jackson}, N.~J., {Browne}, I.~W.~A., {de Bruyn}, A.~G.,
  {Pearson}, T.~J., {Readhead}, A.~C.~S., {Wilkinson}, P.~N., {Biggs}, A.~D.,
  {Blandford}, R.~D., {Fassnacht}, C.~D., {Koopmans}, L.~V.~E., {Marlow},
  D.~R., {McKean}, J.~P., {Norbury}, M.~A., {Phillips}, P.~M., {Rusin}, D.,
  {Shepherd}, M.~C., \& {Sykes}, C.~M. 2003, \mnras, 341, 1

\bibitem[{{Patnaik} {et~al.}(1992){Patnaik}, {Browne}, {Wilkinson}, \&
  {Wrobel}}]{JVAS1}
{Patnaik}, A.~R., {Browne}, I.~W.~A., {Wilkinson}, P.~N., \& {Wrobel}, J.~M.
  1992, \mnras, 254, 655

\bibitem[{{Pearson} {et~al.}(1994){Pearson}, {Shepherd}, {Taylor}, \&
  {Myers}}]{difmap-script}
{Pearson}, T.~J., {Shepherd}, M.~C., {Taylor}, G.~B., \& {Myers}, S.~T. 1994,
  Bulletin of the American Astronomical Society, 26, 1318

\bibitem[{{Petrov} {et~al.}(2005){Petrov}, {Kovalev}, {Fomalont}, \&
  {Gordon}}]{VCS3}
{Petrov}, L., {Kovalev}, Y.~Y., {Fomalont}, E.~B., \& {Gordon}, D. 2005, \aj,
  129, 1163

\bibitem[{{Petrov} {et~al.}(2006){Petrov}, {Kovalev}, {Fomalont}, \&
  {Gordon}}]{VCS4}
---. 2006, \aj, 131, 1872

\bibitem[{{Popov} \& {Kovalev}(1999)}]{PopovKovalev99}
{Popov}, M.~V., \& {Kovalev}, Y.~Y. 1999, Astronomy Reports, 43, 561

\bibitem[{{Shepherd}(1997)}]{difmap}
{Shepherd}, M.~C. 1997, in ASP Conf.\ Ser.\ 125, Astronomical Data Analysis
  Software and Systems VI, ed. G.~{Hunt} \& H.~E. {Payne} (San Francisco: ASP),
  77

\bibitem[{{Verkhodanov} {et~al.}(1997){Verkhodanov}, {Trushkin}, {Andernach},
  \& {Chernenkov}}]{CATS}
{Verkhodanov}, O.~V., {Trushkin}, S.~A., {Andernach}, H., \& {Chernenkov},
  V.~N. 1997, in ASP Conf.\ Ser.\ 125, Astronomical Data Analysis Software and
  Systems VI, ed. G.~{Hunt} \& H.~E. {Payne} (San Francisco: ASP), 322

\bibitem[{{White} {et~al.}(1997){White}, {Becker}, {Helfand}, \&
  {Gregg}}]{FIRST}
{White}, R.~L., {Becker}, R.~H., {Helfand}, D.~J., \& {Gregg}, M.~D. 1997,
  \apj, 475, 479

\bibitem[{{Wilkinson} {et~al.}(1998){Wilkinson}, {Browne}, {Patnaik}, {Wrobel},
  \& {Sorathia}}]{JVAS3}
{Wilkinson}, P.~N., {Browne}, I.~W.~A., {Patnaik}, A.~R., {Wrobel}, J.~M., \&
  {Sorathia}, B. 1998, \mnras, 300, 790

\bibitem[{{Wright} \& {Otrupcek}(1990)}]{pkscat90}
{Wright}, A., \& {Otrupcek}, R. 1990, Parkes Catalog: PKSCAT90 Radio Source
  Catalogue and Sky Atlas (Epping: Australia Telesc. Natl. Facility)

\bibitem[{{Wright} {et~al.}(1994){Wright}, {Griffith}, {Burke}, \&
  {Ekers}}]{Wright_etal94}
{Wright}, A.~E., {Griffith}, M.~R., {Burke}, B.~F., \& {Ekers}, R.~D. 1994,
  \apjs, 91, 111

\bibitem[{{Wright} {et~al.}(1996){Wright}, {Griffith}, {Hunt}, {Troup},
  {Burke}, \& {Ekers}}]{Wright_etal96}
{Wright}, A.~E., {Griffith}, M.~R., {Hunt}, A.~J., {Troup}, E., {Burke}, B.~F.,
  \& {Ekers}, R.~D. 1996, \apjs, 103, 145

\bibitem[{{Wrobel} \& {Ulvestad}(2006)}]{VLBA_summ}
{Wrobel}, J.~M., \& {Ulvestad}, J.~S. 2006, {VLBA status summary,
  http://www.vlba.nrao.edu/astro/obstatus/current/obssum.html}, NRAO

\end{thebibliography}

\end{document}